# Phase diagram of the B–B$_2$O$_3$ system at pressures to 24 GPa

Vladimir Z. Turkevich,[a] Dmitry V. Turkevich[a] and Vladimir L. Solozhenko[b,*]

[a] *Institute for Superhard Materials, National Academy of Sciences of Ukraine, Kiev, 04074 Ukraine*

[b] *LSPM–CNRS, Université Paris Nord, 93430 Villetaneuse, France*

The evolution of topology of the B–B$_2$O$_3$ phase diagram has been studied at pressures up to 24 GPa using models of phenomenological thermodynamics with interaction parameters derived from experimental data on phase equilibria at high pressures and high temperatures.

*Keywords*: B–B$_2$O$_3$ system; high pressure; high temperature; phase diagram.

The B–B$_2$O$_3$ system that includes superhard refractory boron suboxide, B$_6$O [1] (and possibly some novel superhard high-pressure phases [2,3]) has been previously studied at pressure of 5 GPa [4]. However, very recently Solozhenko *et al.* have revised the p-T phase diagram of boron oxide, B$_2$O$_3$ and refined the location of the α-B$_2$O$_3$–β-B$_2$O$_3$–L triple point as 5.4 GPa / 1550 K [5]. This means that at 5 GPa the melting of α-B$_2$O$_3$ takes place, and not of β-B$_2$O$_3$, as previously thought [4,6]. Also the new equilibrium p-T phase diagram of boron [7] should be taken into account. In the present study we performed thermodynamic calculations of high-pressure phase equilibria in the B–B$_2$O$_3$ system using the revised data and constructed phase diagrams at different pressures.

The calculations were carried out using the Thermo-Calc software [8]. Thermodynamic data of phases of the B–B$_2$O$_3$ system at ambient pressure were taken from [4]. The liquid phase was described using the subregular solution model [9], and solid phases – in the framework of the Compound Energy Formalism (CEF) [10]. Pressure dependencies of molar volumes were represented using the Murnaghan approximation [11]. Bulk moduli, their pressure derivatives, and thermal expansion coefficients for B$_6$O and B$_2$O$_3$ were taken from [1,5], and the data on liquid phase, β-B$_{106}$, γ-B$_{28}$, and t′-B$_{52}$, – from [7,12,13]. Parameters of the pressure dependencies of Gibbs energy of the phases in the B–B$_2$O$_3$ system are listed in the Table.

The molar volume of the liquid phase was described by the equation:

$$V_L = V_B x_B + V_{B_2O_3} x_{B_2O_3}$$

Gas phase was considered as ideal gas.

The evolution of the phase diagram of the B–B$_2$O$_3$ system under pressure is shown in the Figure. In addition to the quantitative changes of the diagram parameters (equilibria temperatures, limiting solubilities), changes of the diagram topology are observed i.e. above 8.5 GPa the γ-B$_{28}$ ⇌ t′-B$_{52}$ equilibrium line appears; the congruent type of B$_6$O melting changes to the incongruent one at about 16 GPa, and the L ⇌ β-B$_{106}$ + B$_6$O eutectic changes to the L + t′-B$_{52}$ ⇌ B$_6$O peritectic.

---

[*] vladimir.solozhenko@univ-paris13.fr

Table  Parameters of the pressure dependencies of Gibbs energy

| Phase | $V_0$, m³/mol | B, GPa | B' | $\alpha_0$ | $\alpha_1$ |
|---|---|---|---|---|---|
| α-$B_2O_3$ | 27.25×10⁻⁶ | 40 | 5.5 | 3.0×10⁻⁵ | 1.6×10⁻⁷ |
| β-$B_2O_3$ | 22.38×10⁻⁶ | 170 | 5.0 | 1.5×10⁻⁵ | 2.32×10⁻⁸ |
| $B_6O$ | 31.5×10⁻⁶ | 180 | 6.0 | 1.5×10⁻⁵ | 1.12×10⁻⁸ |
| β-$B_{106}$ | 4.737×10⁻⁶ | 210 | 2.23 | 1.57×10⁻⁶ | 2.64×10⁻⁸ |
| γ-$B_{28}$ | 4.347×10⁻⁶ | 237 | 2.7 | 1.57×10⁻⁶ | 2.64×10⁻⁸ |
| t′-$B_{52}$ | 4.362×10⁻⁶ | 237 | 2.7 | 1.57×10⁻⁶ | 2.64×10⁻⁸ |

$V_0$ is molar volume; B is bulk modulus; $B' = dB/dp$; $\alpha = \alpha_0 + \alpha_1 \times T$ is volume thermal expansion coefficient.

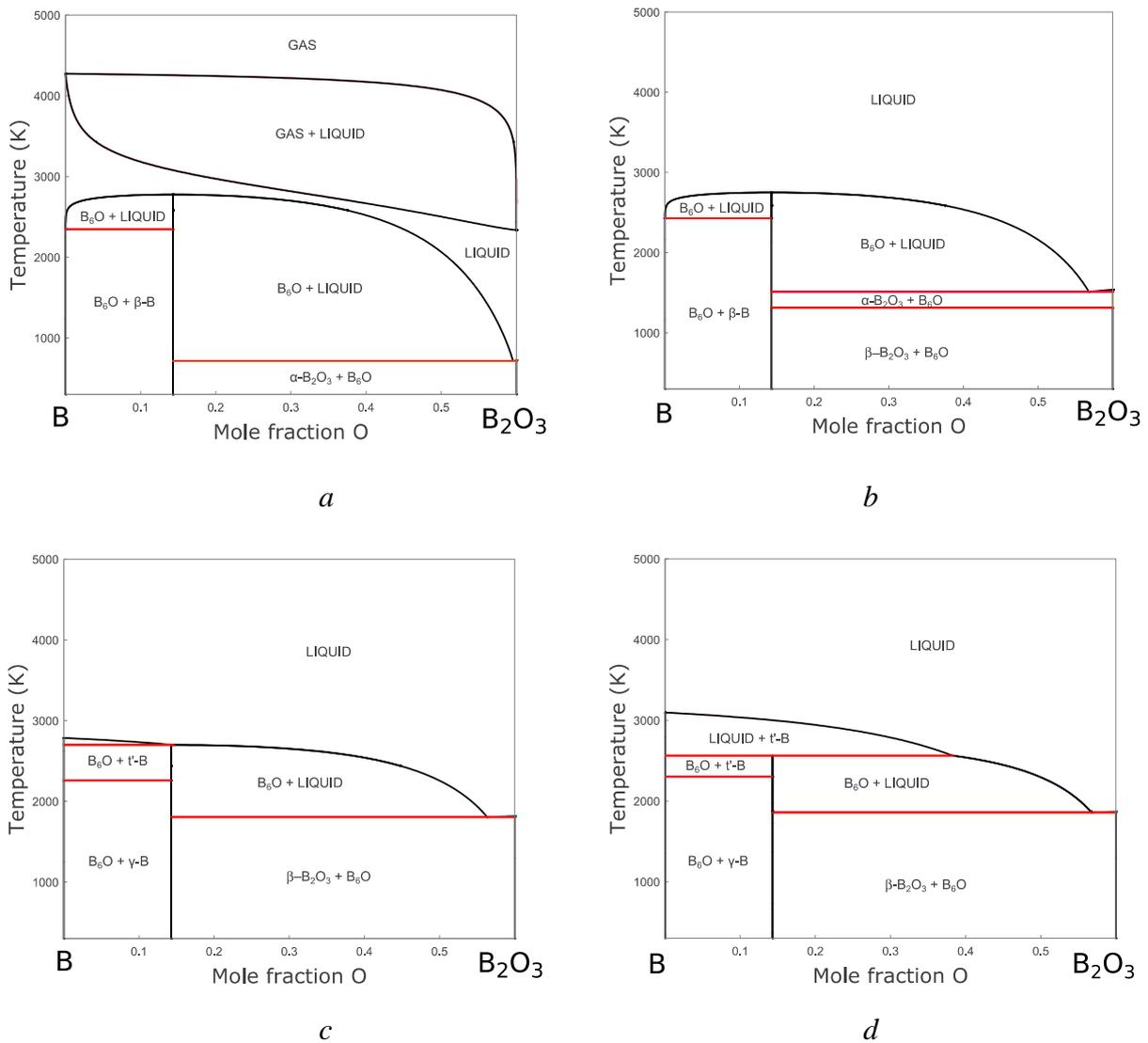

Figure  Phase diagram of the B–$B_2O_3$ system at 0.1 MPa (*a*), 5 GPa (*b*), 16 GPa (*c*) and 24 GPa (*d*); β, γ, t′ are boron allotropes (β-$B_{106}$, γ-$B_{28}$ and t′-$B_{52}$, respectively).

This work was supported by the Agence Nationale de la Recherche (grant ANR-2011-BS08-018). V.Z.T. is grateful to the University Paris Nord for financial support.